\documentstyle[amsmath,amssymb,11pt,twocolumn]{article}
\newcommand{\ket}[1]{\left | #1 \right \rangle}

\begin{document}

\title{\bf Algorithms on Ensemble Quantum Computers\thanks{This work 
was supported in part by grants from the Revolutionary Computing
group at JPL (contract \#961360), and from the DARPA Ultra program
(subcontract from Purdue University \#530--1415--01).}}

\author{P. Oscar Boykin, Tal Mor, Vwani Roychowdhury,
and Farrokh Vatan\thanks{E--mail addresses of the authors are, respectively:
\{boykin, talmo, vwani, vatan\}@ee.ucla.edu.} 
\\ \small Electrical Engineering Department \\ \small UCLA
\\ \small Los Angeles, CA 90095}

\date{ }

\maketitle

\begin{abstract}

In ensemble (or bulk) quantum computation,
measurements of qubits in an individual computer cannot be performed.
Instead, only expectation values can be measured.
As a result of this limitation on the model of computation,
various important algorithms cannot be
processed directly on such computers, and must be modified.
We provide modifications of various existing protocols,  including 
algorithms for universal fault--tolerant computation, Shor's factorization
algorithm (which can be extended to any algorithm 
computing an NP function), and 
some search algorithms to enable
processing them on ensemble quantum computers.

\end{abstract}

\section{Introduction}

Quantum computing is a new type of computing which uses the properties
of quantum mechanics to suggest fast algorithms to several important
problems. 
For example, Shor's algorithm \cite{Shor} for
factoring large numbers is exponentially faster than any known
classical algorithm. Similarly, by utilizing Grover's algorithm \cite{Grover}
it is possible to search a database of size $N$ in time $O(\sqrt{N})$,
instead of $O(N)$ in the classical setting.

NMR computing, first suggested by Cory, Fahmy and Havel~\cite{Cory}, and by
Gershenfeld and Chuang~\cite{CG},
is currently the most promising implementation of quantum
computing, and several quantum algorithms 
involving only few qubits have been demonstrated in the 
labs~\cite{Cory,CG,QEC,Mosca,Teleport2}.
In such NMR systems, each molecule is used as a computer.
Different qubits in the computer are represented by spins of different nuclei.
Many identical molecules (in fact, a macroscopic number) are used in parallel; hence,
this model is called ensemble or bulk quantum computation model. In such bulk models, 
qubits in a single computer cannot be measured, and only expectation
values of a particular bit over all the computers can be read out\footnote{Reading 
the state of $n$ qubits together, as done in many current experiments, is not scalable 
since it requires distinguishing among $2^n$ states.}.

The impossibility of performing
measurements on the individual computers causes severe  
limitations on ensemble quantum computation.
It was generally assumed that rather simple strategies of delaying (or avoiding)
measurements can be used 
to bypass these limitations and to enable the implementation of 
{\em all} quantum algorithms. 
We, however, find that 
for a scalable measurement model such strategies to be 
{\em insufficient} for many algorithms (including Shor's factorization algorithm and 
fault-tolerant computation).

We briefly address other problems related to NMR-computation,
namely, the addressing problem and the pseudo-pure-state (PPS) scaling problem, in Appendix A. 
In the rest of this paper 
we restrict ourselves to issues related solely to the 
{\bf ensemble--measurement problem}. 
While the results here are vital for bulk computation, 
the specific results obtained regarding 
universal and fault-tolerant sets of gates might also be
important for other implementations of quantum 
computing devices where delaying measurements is desired. 

\section{The measurement in ensemble quantum computation}

The measurement process in quantum mechanics 
can be described simply as follows:
To measure the state of a qubit, say
$|\psi \rangle = \alpha|0\rangle+\beta|1\rangle $ in the computation basis
($|0\rangle ; |1\rangle$),
one measures the Hermitian operator (the observable) 
$\sigma_z = \begin{pmatrix} 1 & 0 \\ 0 & -1 \end{pmatrix}$
to get the outcome $\lambda_0 = 1$ with probability $|\alpha|^2$
and $\lambda_1= -1$ with probability
$|\beta|^2$. 
In an NMR ensemble model, the corresponding qubit in every computer
is measured simultaneously, resulting in the expectation
value, i.e., the outcome of
the measurement is a signal of strength proportional to $|\alpha|^2 - |\beta|^2$.  

Clearly, when the outcome of a measurement is expected to be the same on each of the
computers, the ensemble measurement is as good as the standard (single
computer) measurement. 
Usually, this is not the case.
Hence, if the measurement process could yield different
results for different individual computers, 
one would expect that the corresponding algorithm will need modifications
in order to run on an ensemble computer. 

The measurement problem is easily demonstrated in two cases:

{\em Random number generator (RNG):}
Using a single qubit one can easily create 
an RNG. To create a binomial probability 
distribution with parameter $p$ one prepares a state
$\sqrt{p} |0 \rangle + \sqrt{1-p} |1 \rangle$, and measures in the
computational basis to obtain the desired RNG.
This, as far as we know, cannot be done on an ensemble quantum computer, 
where only the expectation value $p \lambda_0 + (1-p) \lambda_1$ can be
classically monitored.
It is unclear yet, whether any algorithm which uses 
an RNG as a subroutine
can still be operated, e.g., using a qubit in a state
$\sqrt{p} |0 \rangle + \sqrt{1-p} |1 \rangle$ to be a control bit of 
the entire process that follows the creation of a random number.

{\em Teleportation:}
Standard teleportation can easily be performed on a three qubit quantum
computer, but strictly speaking, it cannot be performed on an ensemble quantum 
computer. This is because a direct Bell-state measurement of the ensemble quantum 
computer is computationally useless: each computer will yield a random result 
(of the Bell measurement), and on average the outcome is $(1/2) \lambda_0 + (1/2) \lambda_1$
for each of the two measured qubits; hence, there is no way to
decide how to rotate the third qubit in each individual computer.
Yet, a fully-quantum teleportation of the type suggested 
in~\cite{Teleport1} can be, and has been~\cite{Teleport2}, performed on 
ensemble quantum computer:
in this fully-quantum teleportation,
the measurement of an individual computer is never monitored, and a
classically-controlled rotation of the third qubit is replaced by a quantum
control operation, in which the control qubits dephase before being used.

The current algorithms, which have several 
possible measurement outcomes, can 
be sorted into four  groups 
based on the processing which follows the measurement, and the possibility
of avoiding the measurement.
When implemented on ensemble computers, each of the four group requires 
a different adaptation strategy for the algorithm:
\begin{enumerate}
\item For a particular ``desired'' outcome of the entire algorithm,
there is more than one ``good''/``desired'' outcome of an intermediate measurement.
An additional algorithmic 
step is then used to derive the desired final result, and this algorithmic
step can be replaced by a controlled operation
[e.g., error--recovery, in which the final result is the corrected qubit;
Shor's factoring algorithm, in which the same candidate for the ``order''
is obtained from different intermediate measurement outcomes].
\item The algorithm has more than one correct final outcome
and no further processing is done
[e.g., Grover's search algorithm with several solutions] 
\item The algorithm has more than one final result, and some of the results are 
bad/undesired solutions. The algorithm is repeated when bad solution is obtained
[e.g., a wrong factor obtained in Shor's factoring algorithm].
\item The measurement step of the algorithm ought to be replaced with 
available control operations (as in the first case), but such controlled
operation cannot be performed
[e.g., fault--tolerant universal computation].
\end{enumerate}

The first case was recognized before in the seminal work of 
Gershenfeld and Chuang~\cite{CG}. 
When the outcomes of a measurement on various computers are not the same, 
it might be the case that the different measurement outcomes 
can be worked on by a classical algorithm such that a {\em unique} final 
answer is obtained. 
For such algorithms, one can simply delay (or even avoid)
the measurements and 
incorporate the algorithmic step,
which follows the measurement, into the quantum algorithm
(as a controlled operation). 
This 
modified algorithm will now yield a {\em unique} answer on all the
computers. 
It was generally assumed that such strategies of delaying measurements can be used 
to save {\em all} quantum algorithms.
In fact, the strategy's success is restricted only to the cases where the measurements 
can be delayed,
the final outcome is unique, and the final outcome is always obtained\footnote{Although 
we use the term ``delayed'' measurements, in our modified
algorithms sometimes the measurements are not needed at all (and not 
merely delayed).}.
Indeed this strategy works for the error recovery process.
The other cases explained above require 
major modifications of the algorithms.

In the case of algorithms yielding several final good results (case (2)), 
we suggest reordering techniques that provide unique solutions. 
Implementation of search algorithms in the case of multiple solutions on
ensemble computers requires such modifications (derived in Section 4.3).
We note that for a measurement model where all the $2^n$ states of an $n$-qubit
system can be distinguished,
the multiple-solutions case is not a problem. However, as noted earlier, such a scheme
is not practical for any algorithm involving even tens of qubits, and the
exponential resolution requirement makes it no better than a classical computer.

In the case of algorithms having good and bad outcomes (case (3)), we show in Section 4.1
cases where we can solve the
problem by replacing bad results by random data, which do not interfere with the
reading of the good result. 
Previous work by Gershenfeld and Chuang 
\cite{CG} noted that Shor's factorization algorithm 
can be implemented on ensemble quantum computers, by solving the problem as in case
(1). 
However, in addition to problem (1), Shor's algorithm (on ensemble computer)
suffers also from problem (3), and hence the modified algorithm suggested
in \cite{CG} is not sufficient. Hence, the algorithm requires 
a further modification (the randomizing-bad-results strategy) 
in order to work in the general case. 
Alternatively, one might be able to control-repeat the computation in case the 
classical verification
showed that the algorithm yielded a bad output; unfortunately,
such strategy is not easily implementable and cannot be easily
justified; furthermore it leads
to a much longer computation process,
and hence to higher sensitivity to errors.

Case (4) is of pivotal importance to realistic quantum computation. 
The schemes proposed so far for quantum 
fault--tolerant computation usually use an incomplete set of gates, i.e., 
a set of gates that does not generate a dense subset of the group of unitary operations.
In order to complete the set to a universal set, the schemes use interactions
with ancilla qubits, which are then measured
\cite{Shor_fault,KLZ1,Preskill}. Each such measurement
is followed by an application of a unitary operation, $U_j$, 
that depends on the outcome of 
the measurement ($j$). A direct scheme for removing such measurements (followed by
the required unitary operations $U_j$), and replacing them by controlled 
operations, $\Lambda(U_j)$,
will not in general be realizable. This is because,  $\Lambda(U_j)$ might not be realizable
by the incomplete set of fault--tolerant gates. For example, if one attempts
to remove measurements in Shor's scheme for fault--tolerant realization of Toffoli
gate \cite{Shor_fault}, then the corresponding controlled operations would itself
require Toffoli gates! We believe that this issue was not explicitly addressed
in previous works, and we show for the first time how an analysis of error 
propagation and careful design of classical reversible circuits can allow one to
delay measurements in a fault--tolerant manner.

In a prior work, addressing case (4), Aharonov and
Ben--Or \cite{Dorit} have observed that the 
measurements required for fault tolerant computation can be substituted
by reversible classical circuits performing controlled operations. 
In this paper we give an explicit description of this 
process and study in detail
the process of error propagation and how it can be handled in the resulting circuit.
Knill, Laflamme, and Zurek~\cite{KLZ2} followed a different approach  
that potentially does not require measurements.
However, to the best of our knowledge, this approach is incomplete and a proof
of universal fault-tolerant computation is not yet available.
For example, a measurement-free implementation of the Hadamard gate using that approach 
has not been demonstrated.

Finally, Peres~\cite{Peres} also 
discusses the possibility of measurement--free
encoding and decoding procedures in quantum error--correction.
However, in his scheme the quantum information is transformed to
a single qubit, while we suggest a method that is suitable 
for fault--tolerant computation.

\section{Obtaining a universal and fault--tolerant set of gates}

The idea of quantum fault--tolerant
computation~\cite{Shor_fault,Dorit,KLZ1,Kitaev,Preskill} 
can be described briefly as follows.
Suppose that we have a noisy quantum circuit $C$ which we 
want to simulate by
a fault--tolerant circuit $\widetilde C$. 
In one level of such a circuit, the regular bits are replaced by logical
bits $\ket{0}_L$ and $\ket{1}_L$, where these are some entangled
states of a block of 
physical qubits. While $C$ operates on data qubits, in the
circuit $\widetilde C$ all operations are performed on encoded data, 
i.e., each data
qubit or a set of data qubits is represented as a block of qubits that belongs
to some quantum error--correcting code. Then each operation of $C$ performed
by a gate $g_j$ is simulated by a procedure (subcircuit) 
$\widetilde{g_j}$ in the circuit $\widetilde C$ such that
in $\widetilde{g_j}$ each computation transforms codewords to codewords.
In order to avoid accumulation of errors, after each computation in $\widetilde{g_j}$ 
a ``correction procedure'' is performed to correct any error that is introduced
in that computation. So in the fault--tolerant circuit $\widetilde C$ each computation
step is followed by a correction step.

The 
operations on the encoded qubits introduce a large number of additional gates and qubits, and
unless one is careful, it is possible that more errors are introduced than can
be corrected by the code. 
To avoid any such catastrophic accumulation of errors, it is desirable that 
the operations in the fault-tolerant
circuits prevent ``spreading of errors'' by making sure that each gate error causes
a single error in each block. It is useful now to review how errors propagate in quantum circuits.
For example, consider the CNOT (controlled--not) gate which 
performs the operation
$\ket{a}_c\ket{b}_t \mapsto \ket{a}_c\ket{a\oplus b}_t$
in the computation basis; for the rest of this paper, 
we shall drop the subscripts $c$ (control) and $t$ (target)  and designate the 
control bit as the one on the left side. Clearly, applying CNOT operation
from one bit to many target bits can propagate one bit error from the control bit to all the
target bits. On the other hand, applying CNOT from many control bits to one
target bit can propagate one phase error from the target bit to all the control bits.
It is easy to observe this ``back'' propagation of the phase errors: 
if we apply CNOT on the state $(\ket{0}+\ket{1})\otimes
(\ket{0}+\ket{1})$ and there is a phase error in the target qubit, we will get
\begin{gather*}
    \ket{0}\otimes(\ket{0}-\ket{1})+\ket{1}\otimes(\ket{1}-\ket{0})= \hspace{1.2cm} \\
      \hspace{2cm} (\ket{0}-\ket{1})\otimes(\ket{0}-\ket{1}) 
\end{gather*}
which results in a phase error in the control qubit. Hence, fault-tolerant computation requires that 
this gate be applied only in the case where the control qubit $\ket{a}$ and the 
target qubit $\ket{b}$ belong to different blocks. Furthermore, this error-propagation
phenomenon is also true for other controlled operations, and this  
 motivated a {\em sufficient} condition for fault tolerance:
only perform bitwise operations  or transversal operations on qubits 
within a code.  It is,  however,  
{\em not} a {\em necessary}
condition for fault--tolerance, and 
careful constructions may allow one to apply control gates 
from many control bits onto one target bit, without destroying the fault-tolerant
computation, to resolve the catch-22 problem we observe in the following discussions.

To get a quantum fault--tolerant computation, it is enough to show that for a
{\em universal} set of quantum gates  the above mentioned procedure on
the encoded data is possible. Quantum fault--tolerant schemes usually
(see, e.g., \cite{Shor_fault,Preskill}) depend on measurements to 
ensure that the set of the operations permissible on encoded data (i.e.,
codewords in a quantum error--correcting code) is actually a universal set.
Some of the gates in the universal set do not require measurements, e.g., the
operations $H$, $\sigma_z^{1/2}$, and CNOT. 
[For CSS codes~\cite{Shor_fault}, 
each of these logical gates can simply be achieved by performing the same gate bit-wise
on the individual qubits
(e.g., $H$ is achieved on code words via applying $H$ on individual qubits), 
but the bit-wise $\sigma_z^{1/2}$ yield a  
$\sigma_z^{-1/2}$ logical gate, hence  
requires an additional step of bit-wise $\sigma_z$, to yield the desired logical gate.]
In existing suggestions (except~\cite{KLZ2} as previously explained), 
at least one gate  (e.g., Toffoli in \cite{Shor_fault} and $\sigma_z^{1/4}$ 
in~\cite{sigma_z}) requires measurements.

There is always a simple scheme that potentially allows 
one to postpone measurements of ancilla qubits in 
quantum computation.
Recall that a measurement is followed by an operation $U_j$, which 
is a unitary operation performed on the data based on the outcome of a measurement
on the ancilla qubits (and $U_j$ can be performed fault-tolerantly using the given,
non-universal, set of operations). 
As explained in Section 2, the scheme for delaying the measurement can 
be successfully implemented only if the controlled operations $\Lambda(U_j)$'s 
are in the set of
{\em available} measurement--free operations; i.e., these control operations can be
implemented on encoded data fault--tolerantly and directly without using 
any measurements. However, in  
the cases investigated
so far, it is not the case that the required controlled operations 
$\Lambda(U_j)$ are implementable in a direct fault-tolerant manner.
For instance, in Shor's fault--tolerant set of gates \cite{Shor_fault}, a
measurement is required for the preparation of a Toffoli gate, but a
Toffoli gate is required if we want to delay that measurement.
This is because the measurement is followed by a controlled--NOT operation, and
hence can only be replaced by a controlled--controlled--NOT which is a Toffoli gate.
This seems like 
a catch-22 situation! 
\footnote{Similarly, 
in the fault-tolerant universal set of gates suggested in~\cite{sigma_z},
the generation of the $\sigma_z^{1/4}$ gate without measurements leads to a catch-22
problem; 
a $\sigma_z^{1/2}$ gate (which follows the measurement) 
need to be replaced by a $\Lambda(\sigma_z^{1/2})$ gate, 
which is not available as long as the $\sigma_z^{1/4}$ gate is not available.}

However, the solution comes from the vital observation
that some operations 
need protection only from the bit errors, 
and do not need to use full quantum codes.

By replacing the ``quantum ancilla'' (in a logical basis $|0\rangle_L$ and
$|1\rangle_L$) by a ``classical ancilla'' in a ``classical'' basis
$|\vec{0}\,\rangle=\ket{0\cdots0}$ and $|\vec{1}\,\rangle=\ket{1\cdots1}$, 
we can use the classical ancilla to perform 
$\Lambda(U_j)$ in a fault-tolerant manner,
and this can be done in the two cases where the outcomes are
the Toffoli gate required for the Shor's basis,
and the $\sigma_z^{1/4}$ gate required for the basis of~\cite{sigma_z}.
One can interpret the classical basis as the classical repetition code.
We call the ancilla in these states ``classical'' since a classical error-correction
code can be used to correct bit errors in it. 
Clearly, phase errors are not corrected
in the classical ancilla, yet we found that the use of such a classical 
ancilla is still good enough for
our purpose.

\vspace{6mm}
{\bf Replacing Measurements of Encoded Ancilla Qubits:}\\

In the following we shall replace the measurement of the quantum ancilla followed by
the operation $U$ acting on the quantum data, by a sequence
of operation: we copy the two basis states of a quantum ancilla into a classical
ancilla, we perform classical error correction on the classical ancilla, and we use 
the classical ancilla 
as a control bit for performing the operation
$\Lambda(U_j)$ with the quantum data as the target bit.

The measurement of the quantum ancilla in the original protocol
is done as follows 
\cite{Preskill}:  measure each of the physical qubits, 
and perform a classical error correction 
on the outcome of this measurement to determine the state of the ancilla.  For example, if
the 7-bit CSS code \cite{Shor_fault} is used to encode data, then a 
measurement will yield a possibly corrupted codeword of a classical 7-bit Hamming code. After
classical error correction, if the parity of the codeword is ``even'' then the ancilla has
collapsed to the state $\ket{0}_L$, otherwise to the state $\ket{1}_L$. 
Classical error correction is enough because phase errors before 
a measurement will not change the outcome probabilities.



As a first step toward removing such a measurement, we propose a new gate that copies
an encoded quantum ancilla word onto a classical ancilla:
\begin{equation}
{\cal N}:\left\{
\begin{array}{rcl}
\ket{0}_L\otimes|\vec{0}\,\rangle & \longrightarrow & \ket{0}_L\otimes|\vec{0}\,\rangle,\\
\ket{0}_L\otimes|\vec{1}\,\rangle & \longrightarrow & \ket{0}_L\otimes|\vec{1}\,\rangle,\\
\ket{1}_L\otimes|\vec{0}\,\rangle & \longrightarrow & \ket{1}_L\otimes|\vec{1}\,\rangle,\\
\ket{1}_L\otimes|\vec{1}\,\rangle & \longrightarrow & \ket{1}_L\otimes|\vec{0}\,\rangle. 
\end{array} \right .
\label{N}
\end{equation}
Let $\cal N$ be a unitary operation that implements the above transformation.
[We show in the next subsection that 
this operation can be done fault--tolerantly.]

\begin{figure*} \begin{center} \unitlength=.38mm 
\begin{picture}(330,140)(0,0) 
\put(0,0){\line(1,0){330}}\put(0,30){\line(1,0){330}}\put(0,40){\line(1,0){330}}
\put(0,50){\line(1,0){330}}\put(0,80){\line(1,0){330}}\put(0,90){\line(1,0){330}}
\put(0,100){\line(1,0){330}}\put(0,110){\line(1,0){330}}\put(0,120){\line(1,0){330}}
\put(0,130){\line(1,0){330}}\put(0,140){\line(1,0){330}}
\multiput(20,140)(10,-10){7}{\circle*{5}}
\multiput(20,0)(10,0){7}{\circle{8}}
\put(20,140){\line(0,-1){144}}\put(30,130){\line(0,-1){134}}
\put(40,120){\line(0,-1){124}}\put(50,110){\line(0,-1){114}}
\put(60,100){\line(0,-1){104}}\put(70,90){\line(0,-1){94}}
\put(80,80){\line(0,-1){84}}
\multiput(110,140)(10,-20){4}{\circle*{5}}
\multiput(110,50)(10,0){4}{\circle{8}}
\put(110,140){\line(0,-1){94}}\put(120,120){\line(0,-1){74}}
\put(130,100){\line(0,-1){54}}\put(140,80){\line(0,-1){34}}
\multiput(170,110)(10,-10){4}{\circle*{5}}
\multiput(170,40)(10,0){4}{\circle{8}}
\put(170,110){\line(0,-1){74}}\put(180,100){\line(0,-1){64}}
\put(190,90){\line(0,-1){54}}\put(200,80){\line(0,-1){44}}
\multiput(230,30)(10,0){4}{\circle{8}}
\put(230,130){\circle*{5}}\put(240,120){\circle*{5}}
\put(250,90){\circle*{5}}\put(260,80){\circle*{5}}
\put(230,130){\line(0,-1){104}}\put(240,120){\line(0,-1){94}}
\put(250,90){\line(0,-1){64}}\put(260,80){\line(0,-1){54}}
\multiput(290,30)(0,10){3}{\circle{8}}
\multiput(290,34)(0,10){3}{\line(0,-1){8}}
\multiput(290,0)(10,0){2}{\circle{8}}
\put(290,4){\line(0,-1){8}}
\multiput(300,50)(0,-10){3}{\circle*{5}}
\put(300,50){\line(0,-1){54}}
\multiput(310,30)(0,10){3}{\circle{8}}
\multiput(310,34)(0,10){3}{\line(0,-1){8}}
\put(0,110){\makebox(0,0){$\left\{\begin{tabular}{c}\vspace{21mm}\end{tabular}\right.$}}
\put(-30,110){\makebox(0,0){\small codeword}}
\put(-7,0){\makebox(0,0){\small$\ket{0}$}}
\put(-7,30){\makebox(0,0){\small$\ket{0}$}}
\put(-7,40){\makebox(0,0){\small$\ket{0}$}}
\put(-7,50){\makebox(0,0){\small$\ket{0}$}}
\put(328,40){\makebox(0,0){$\left.\begin{tabular}{c}\vspace{7mm}\end{tabular}\right\}$}}
\put(358,40){\makebox(0,0){\footnotesize $\ket{\mathrm\sf syndrome}$}}
\put(340,0){\makebox(0,0){\small$\ket{b}$}}

\end{picture} \end{center} 
\caption{The operation ${\cal N}_1$. Note that the circuit shows the generation of only one
classical target bit $\ket{b}$; the operations on the last bit 
have to be repeated to generate multiple target bits. } 
\label{N_fig} \end{figure*}
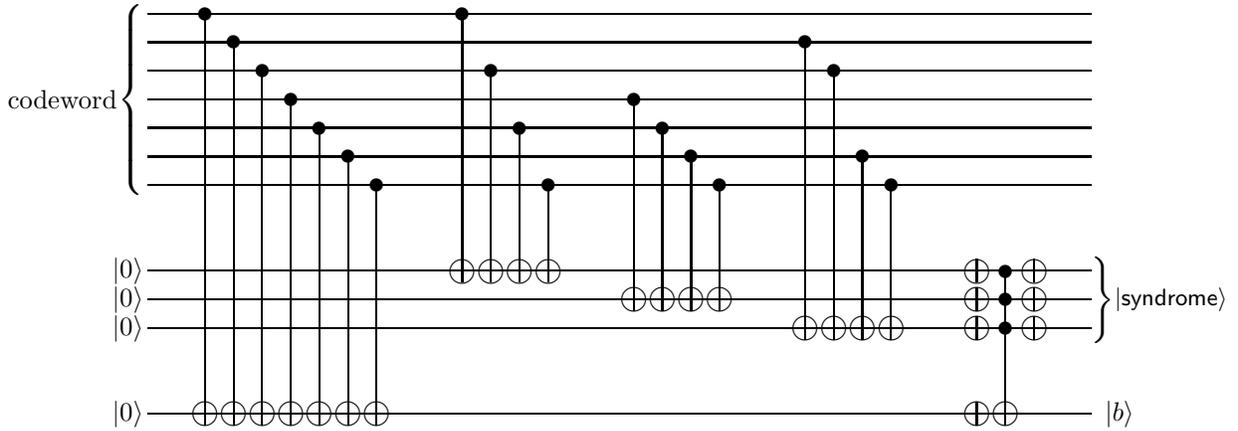

With this operation ($\cal N$), 
the quantum bit is ``copied'' onto the classical ancilla.  
Since the repetition code can only correct for bit errors in the classical ancilla,
one must make sure that the classical ancilla can still be used to perform 
$\Lambda(U_j)$ without putting the quantum data in jeopardy.  
This, however, is not a problem, since phase errors are transmitted from 
{\em target} bit to control bit, hence cannot be transmitted from the
classical ancilla (control) to the quantum data (target).
This leads to the most interesting 
and possibly counter intuitive aspect of our scheme: the data in the classical 
repetition code, or any classical function of this data, {\em can} act as control bits in a 
bitwise controlled-$U$ operation onto quantum data.

We shall show later two cases where indeed the operations between the classical
ancilla and the quantum data can be performed bit-wise while the same
operations cannot be performed bit-wise between quantum ancilla and the quantum
data (as the naive solution of delaying measurements would have suggested).

Note that the quantum data may add phase errors to the 
repetition code, but that is of no concern to us, since 
also in the ``measured'' case, the classical 
repetition code has lost phase coherence.  If there are $t$ bit errors 
in the repetition code, it will result in $t$ errors in the quantum data.  Fortunately, 
bit errors {\em are} corrected in the repetition code. Hence, 
{\em the operation $\cal N$
enables one to create universal bases without measurement}.

\vspace{6mm}
{\bf The operation $\cal N$: quantum-to-classical controlled--NOT.}
In Figure \ref{N_fig}, we 
represent a circuit that computes operation ${\cal N}_1$ for the seven--bit CSS code,
where ${\cal N}_1$ stands for Eq.(\ref{N}) with only one bit of the classical ancilla.
The syndrome ancilla bits are used to prevent the spread of one
bit error from the quantum ancilla into the classical bit. Only two errors
(in any of the inputs, the gates or the time steps) shall yield an error
in the classical bit.

This is not the complete circuit; in the complete circuit, the same computation
on the bottom four bits is repeated $n$ times, where $n$ is the number of qubits in a
codeword.  At each repetition
stage the syndrome bits are discarded, and another bit
$b_i$ is created ($1\leq i\leq n$).  
In principle, the syndrome bits could be ignored,
reset, or measured.  These bits will not effect the operation beyond
their use as a form of error detection in the codeword.
The bits $b_i$ are then corrected (to yield the classical $0$ or $1$) using a majority
vote.

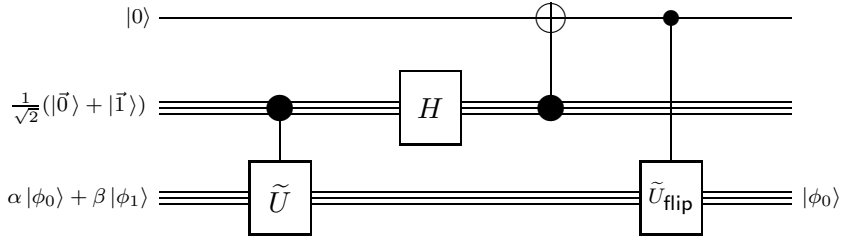
\begin{figure*} \begin{center} \unitlength=.4mm 
\begin{picture}(250,65)(-30,0) 
\put(0,0){\line(1,0){30}}\put(0,2){\line(1,0){30}}\put(0,-2){\line(1,0){30}}
\put(40,0){\makebox(0,0){\framebox(20,24){$\widetilde{U}$}}}
\put(0,30){\line(1,0){80}}\put(0,32){\line(1,0){80}}\put(0,28){\line(1,0){80}}
\put(90,30){\makebox(0,0){\framebox(20,24){$H$}}}
\put(40,30){\circle*{9}}\put(40,30){\line(0,-1){18}}
\put(50,0){\line(1,0){110}}\put(50,2){\line(1,0){110}}\put(50,-2){\line(1,0){110}}
\put(100,30){\line(1,0){110}}\put(100,32){\line(1,0){110}}\put(100,28){\line(1,0){110}}
\put(170,0){\makebox(0,0){\framebox(20,24){\scriptsize $\widetilde{U}_{\mbox{\sf flip}}$}}}
\put(180,0){\line(1,0){30}}\put(180,2){\line(1,0){30}}\put(180,-2){\line(1,0){30}}
\put(0,60){\line(1,0){210}}
\put(130,30){\circle*{9}}\put(130,30){\line(0,1){35}}
\put(130,60){\circle{10}}
\put(170,60){\circle*{5}}\put(170,60){\line(0,-1){48}}
\put(-27,0){\makebox(0,0){\scriptsize$\alpha\ket{\phi_0}+\beta\ket{\phi_1}$}}
\put(-27,30){\makebox(0,0){\scriptsize$\frac{1}{\sqrt{2}}(|\vec{0}\,\rangle+|\vec{1}\,\rangle$)}}
\put(-7,60){\makebox(0,0){\scriptsize$\ket{0}$}}
\put(220,0){\makebox(0,0){\scriptsize$\ket{\phi_0}$}}

\end{picture} \end{center} 
\caption{Preparing an eigenvector. } 
\label{eigen_fig} \end{figure*}

\begin{figure*} \begin{center} \unitlength=.3mm 
\begin{picture}(260,70)(0,0) 
\put(0,0){\line(1,0){60}}\put(0,30){\line(1,0){60}}
\put(0,60){\line(1,0){110}}
\put(70,15){\makebox(0,0){\framebox(20,54){${\cal N}$}}}
\put(30,60){\circle*{6}}\put(30,60){\line(0,-1){35}}
\put(30,30){\circle{10}}
\put(125,60){\makebox(0,0){\framebox(30,24){\small${\sigma_z}^{\frac{1}{2}}$}}}
\put(80,0){\line(1,0){100}}\put(80,30){\line(1,0){100}}
\put(140,60){\line(1,0){40}}
\put(125,0){\circle*{6}}\put(125,0){\line(0,1){48}}
\put(-12,0){\makebox(0,0){\small$|\vec{0}\,\rangle$}}
\put(-12,30){\makebox(0,0){\small$\ket{\psi_0}$}}
\put(-12,60){\makebox(0,0){\small$\ket{x}_L$}}
\put(210,60){\makebox(0,0){\small${\sigma_z}^{1/4}\ket{x}_L$}}
\put(180,15){\makebox(0,0){$\left.\begin{tabular}{c}\vspace{10mm}\end{tabular}\right\}$}}
\put(250,15){\makebox(0,0){\scriptsize$\frac{1}{\sqrt{2}}(\ket{0}_L|\vec{0}\,\rangle
                    +e^{\frac{i\pi}{4}}\ket{1}_L|\vec{1}\,\rangle)$}}
\end{picture} \end{center} 
\caption{Fault--tolerant ${\sigma_z}^{1/4}$ without measurement. } 
\label{sigma_fig} \end{figure*}
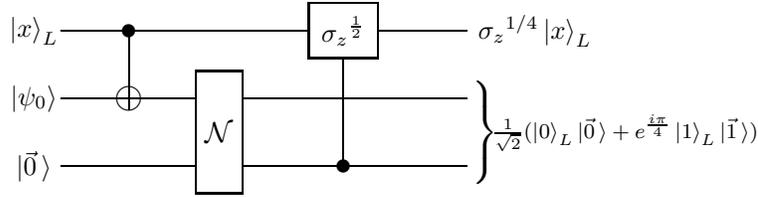
The circuit ${\cal N}_1$ flips the bit $b$ if the quantum ancilla (acting here
as a control bit) is $\ket{1}_L$, and does nothing otherwise.
This circuit operates properly 
as long as there is up to one bit error in the quantum data 
(there can actually be an unlimited 
number of phase errors).  
Note that phase errors in the lower part will spread to the quantum ancilla; however,  
this is of no consequence, since the quantum ancilla never interact with 
the quantum data in later stages.
Bit errors in the quantum ancilla are important, since the process is repeated
$n$ times, hence bit errors, created in the quantum ancilla 
at initial stage of ${\cal N}_1$, will spread errors into the next bits
of the classical
ancilla. Fortunately, 
bit errors are not transmitted from the classical to quantum section,
and the quantum ancilla cannot be disturbed by a bit error in bits of the classical
ancilla or the syndrome ancilla.  If there is one 
error in the $|0\rangle$ bits used to store the syndrome it will cause an error in the 
single classical bit.  But such errors must be overcome by repeating this circuit $n$ 
times with fresh syndrome bits for each repetition.  
At that point we will have a repetition code that will successfully recover 
from $k'$ errors.  Once this number $k'$ is equal to, 
or greater than, the number of errors, $k$, that 
the \emph{quantum} code can correct for, we may stop.
For a probability $p$ of an error (per gate, per input bit, and per 
delay line) the resulting error rate of this circuit is $O(p^2)$,
as required for fault tolerant computation.
The threshold can easily be calculated by counting the potential places for two errors, 
and the threshold can be much improved by 
enhancing the parallelism, and by repeating ${\cal N}_1$ only $2k+1$ times
(e.g., with the 7-bit quantum code, that is $n=7$, which corrects $k=1$ error,
it is enough to repeat the circuit $3$ times, correct the outcome using a majority vote,
and then copy the result into seven bits).

Any required classical reversible fault--tolerant 
calculation can be performed on the classical ancilla.
Finally, it is used 
as control bits in bitwise operations back onto the quantum data.

\vspace{6mm}
{\bf Creating the special states required for fault--tolerant universal
computation, without using a measurement}


Our method is general and can be described as follows. Assume that a quantum
code of length $n$ is used for encoding data. Suppose that $U\in\mbox{\bf U}(2^l)$ 
[for our purpose it is enough to consider up to three qubits ($l=3$) operations],
and
$\widetilde{U}=U^{\otimes n}$ is the unitary operation on the codewords obtained
by applying $U$ bitwise. Suppose that $\widetilde{U}$ has
eigenvectors $\ket{\phi_0}$ and $\ket{\phi_1}$ such that
\[   \widetilde{U}\ket{\phi_0}=\ket{\phi_0} \qquad \mathrm{and} \qquad 
          \widetilde{U}\ket{\phi_1}=-\ket{\phi_1} . \]
Then the quantum circuit in Figure \ref{eigen_fig} outputs the eigenvector $\ket{\phi_0}$
if the input state is 
$\alpha\ket{\phi_0}+\beta\ket{\phi_1}$ 
(for any $\alpha$, $\beta$).
In this figure $\widetilde{U}_{\mathrm{\sf flip}}$ is a unitary operation that maps
$\ket{\phi_0}$ on $\ket{\phi_1}$ and vice versa.
The operations $\Lambda(\widetilde{U})$ (i.e.,
the controlled--$\widetilde{U}$), and $H$
are applied bitwise.  
The last two controlled operations will be explained in the sequel.

This scheme is practical if it is possible to prepare a state 
$\alpha\ket{\phi_0}+\beta\ket{\phi_1}$, where it does not matter what is the values of 
$\alpha$ and $\beta$. In this circuit the first line is a single parity bit,
each of the second and third inputs is a block 
of $n$ qubits, containing the cat-states lines and the special state lines respectively. 
The third gate, the controlled-not gate which we call here $P$, 
is a Parity gate which calculates the parity
of the cat-state lines and puts the result in the parity bit.  
It is done by a sequence of controlled-not from each control bit onto one target bit.
The figure only demonstrates the creation of one parity bit 
$\ket{\phi_0}$ in an unprotected manner (as far as a bit error in the parity bit
is of concern).
The real circuit is a bit different:
The operations $\Lambda(\widetilde{U})$, $H$ and $P$, are repeated $n$ times,
each time with fresh cat-states and fresh parity bit (but on the same 
special state's lines).
Then a majority vote is calculated on the parity bits, 
in order to reduce the probability
that an error in a cat state or in the parity bit will ruin the result.  
Then the i$n$ parity results are corrected, so that the probability of two errors
becomes low [that is, of order $O(p^2)$]. Finally, the parity result
is used to control 
$\widetilde{U}_{\mathrm{\sf flip}}$ 
in a bit-wise manner, so that the special state is created via a fault tolerant
operation.


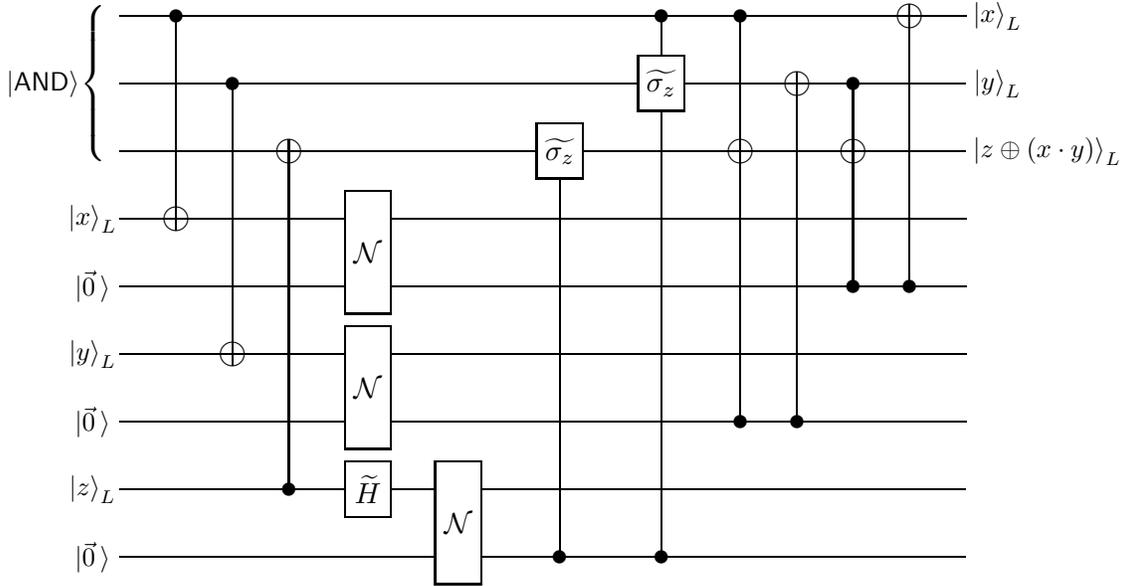
\begin{figure*} \begin{center} \unitlength=.3mm 
\begin{picture}(380,240)(0,0) 

\put(0,0){\line(1,0){140}}\put(0,30){\line(1,0){100}}\put(120,30){\line(1,0){20}}
\put(110,30){\makebox(0,0){\framebox(20,24){$\widetilde{H}$}}}
\put(150,15){\makebox(0,0){\framebox(20,54){${\cal N}$}}}
\put(110,75){\makebox(0,0){\framebox(20,54){${\cal N}$}}}
\put(110,135){\makebox(0,0){\framebox(20,54){${\cal N}$}}}
\put(0,60){\line(1,0){100}}\put(0,90){\line(1,0){100}}
\put(0,120){\line(1,0){100}}\put(0,150){\line(1,0){100}}
\put(195,180){\makebox(0,0){\framebox(20,24){$\widetilde{\sigma_z}$}}}
\put(240,210){\makebox(0,0){\framebox(20,24){$\widetilde{\sigma_z}$}}}
\put(0,180){\line(1,0){185}}\put(0,210){\line(1,0){230}}
\put(250,210){\line(1,0){125}}\put(205,180){\line(1,0){170}}
\put(120,150){\line(1,0){255}}\put(120,120){\line(1,0){255}}
\put(120,90){\line(1,0){255}}\put(120,60){\line(1,0){255}}
\put(160,30){\line(1,0){215}}\put(160,0){\line(1,0){215}}\put(0,240){\line(1,0){375}}
\put(25,240){\circle*{6}}\put(25,240){\line(0,-1){95}}
\put(25,150){\circle{10}}
\put(50,210){\circle*{6}}\put(50,210){\line(0,-1){125}}
\put(50,90){\circle{10}}
\put(75,30){\circle*{6}}\put(75,30){\line(0,1){155}}
\put(75,180){\circle{10}}
\put(195,0){\circle*{6}}\put(195,0){\line(0,1){168}}
\put(240,0){\circle*{6}}\put(240,0){\line(0,1){198}}
\put(240,240){\circle*{6}}\put(240,240){\line(0,-1){18}}
\put(275,240){\circle*{6}}\put(275,240){\line(0,-1){180}}
\put(275,180){\circle{10}}\put(275,60){\circle*{6}}
\put(300,60){\circle*{6}}\put(300,60){\line(0,1){155}}
\put(300,210){\circle{10}}
\put(325,210){\circle*{6}}\put(325,210){\line(0,-1){90}}
\put(325,180){\circle{10}}\put(325,120){\circle*{6}}
\put(350,120){\circle*{6}}\put(350,120){\line(0,1){125}}
\put(350,240){\circle{10}}
\put(-12,0){\makebox(0,0){\small$|\vec{0}\,\rangle$}}
\put(-12,30){\makebox(0,0){\small$\ket{z}_L$}}
\put(-12,60){\makebox(0,0){\small$|\vec{0}\,\rangle$}}
\put(-12,90){\makebox(0,0){\small$\ket{y}_L$}}
\put(-12,120){\makebox(0,0){\small$|\vec{0}\,\rangle$}}
\put(-12,150){\makebox(0,0){\small$\ket{x}_L$}}
\put(-4,210){\makebox(0,0){$\left\{\begin{tabular}{c}\vspace{17mm}\end{tabular}\right.$}}
\put(-34,210){\makebox(0,0){\small$\ket{\mathrm{\sf AND}}$}}
\put(389,240){\makebox(0,0){\small$\ket{x}_L$}}
\put(389,210){\makebox(0,0){\small$\ket{y}_L$}}
\put(411,180){\makebox(0,0){\small$\ket{z\oplus(x \cdot y)}_L$}}

\end{picture} \end{center} 
\caption{Fault--tolerant Toffoli without measurement.} 
\label{toffoli_fig} \end{figure*}

\vspace{6mm}
{\bf Fault--tolerant ${\sigma_z}^{1/4}$ without measurement.}

Let $\cal B$ be the basis consisting of $H$ (Hadamard), ${\sigma_z}^{1/2}$, and
CNOT. The operations in $\cal B$ are fault--tolerant, 
simply because they can be 
applied to encoded data bitwise (when standard codes are used).
But $\cal B$ is not universal. One way to make
$\cal B$ universal is to add the Toffoli gate to it. Another way is to add the gate
${\sigma_z}^{1/4}$, as shown in \cite{sigma_z}. 
The advantages of this latter set
of gates are that it is 
(a) simple to be implemented, 
(b) simple to be proven universal,
and
(c) simple to operate with delayed measurements.

We show here how it is possible to implement this operation on codewords without
using any measurement. 
This scheme is a modified version of the original method for
implementing ${\sigma_z}^{1/4}$ on codewords \cite{sigma_z}, and it does
not use measurements. 

First, we need to prepare the following state
\[ \ket{\psi_0}=\frac{1}{\sqrt{2}}\left(\ket{0}_L+e^{\frac{i\pi}{4}}\ket{1}_L\right).\]
This state can be prepared with a circuit of form given in Figure \ref{eigen_fig}. 
For this purpose, let $U=e^{\frac{i\pi}{4}}\sigma_x\sigma_z{\sigma_z}^{1/2}$ 
and $\ket{\psi_1}=\frac{1}{\sqrt{2}}\left(\ket{0}_L-e^{\frac{i\pi}{4}}\ket{1}_L\right)$. 
Then $\widetilde{U}\ket{\psi_0}=\ket{\psi_0}$, 
$\widetilde{U}\ket{\psi_1}=-\ket{\psi_1}$, and $U_{\mathrm{\sf flip}}=\sigma_z$.

Now we are ready to describe the fault--tolerant ${\sigma_z}^{1/4}$
without measurement. 
Then the circuit in Figure \ref{sigma_fig} shows 
the fault--tolerant implementation 
of ${\sigma_z}^{1/4}$ on a codeword $\ket{x}_L$. 
In this circuit, $\cal N$ is the 
unitary operation defined in (\ref{N}). 
Apart 
from  replacing the standard
measurements by the $\cal N$ circuit, this figure is exactly the same as
the one drawn in~\cite{sigma_z} to implement the ${\sigma_z}^{1/4}$ gate.
In this figure 
each input in fact denotes a block of qubits, and operations are bitwise.

\vspace{6mm}
{\bf Fault--tolerant Toffoli without measurement.}

The more conventional (and more complicated) set of universal
fault-tolerant gates contain the Toffoli instead of the
${\sigma_z}^{1/4}$.  

We show explicitly how to implement Toffoli on encoded data without using any 
measurement. This scheme is a modified version of Shor's original method for
implementing Toffoli on codewords \cite{Shor_fault}. 
The method is similar to the one applied to ${\sigma_z}^{1/4}$.

In Shor's method (as in the other bases we have shown before) a preparation
of a special state is required, hence we first
prepare the state
\begin{equation}
\mbox{\footnotesize
  $\ket{\mathrm{\sf AND}}={\textstyle \frac{1}{2}}\left (
              \ket{000}_L+\ket{010}_L+\ket{100}_L+\ket{111}_L\right ) ,$ }
\label{AND}
\end{equation}
without using measurement, based on our ``creating a special state''
technique.

To get $\ket{\mathrm{\sf AND}}$
we let $U=\Lambda(\sigma_z)\otimes\sigma_z$, and we chose
\[ \mbox{\small$\ket{\,\overline{\mathrm{\sf AND}}\,}={\textstyle \frac{1}{2}}\left ( 
               \ket{001}_L+\ket{011}_L+\ket{101}_L+\ket{110}_L\right )$}. \]
Then $\widetilde U\ket{\mathrm{\sf AND}}=\ket{\mathrm{\sf AND}}$,
$\widetilde U\ket{\,\overline{\mathrm{\sf AND}}\,}=-\ket{\,\overline{\mathrm{\sf AND}}\,}$,
$U_{\mathrm{\sf flip}}=I\otimes I\otimes \sigma_x$, and
\[ {\textstyle \frac{1}{\sqrt{2}}}\left ( \ket{\mathrm{\sf AND}} +
    \ket{\,\overline{\mathrm{\sf AND}}\,}\right ) = 
     (\widetilde H\otimes\widetilde  H\otimes\widetilde  H)\ket{000}_L .\]
A different solution to this step was given (independently) by D.~Aharonov and
M.~Ben-Or~\cite{Dorit2}.

Now we are ready to describe the fault--tolerant Toffoli without measurement. 
This
procedure is presented in Figure \ref{toffoli_fig}. 
In this circuit $\cal N$ is the 
unitary operation defined in (\ref{N}); 
apart for replacing the standard
measurements by our $\cal N$ circuit, this figure is exactly the same as
the one drawn by Preskill~\cite{Preskill} to describe Shor's way of obtaining
the Toffoli gate.

Note that in this figure each input
represents a block of qubits and operations on these blocks are defined in
the natural way. Also note that the first three top outputs of this circuit are in a
tensor product with the rest of the outputs.

\section{Quantum algorithms}

Here we study different known quantum algorithms that
cannot be implemented directly on ensemble quantum computers and we provide modifications
to make them suitable for such computers.

\subsection{The factorization algorithm}

In the Shor's factorization algorithm the aim is to factor a large number
$n$. To do so, one uses
a random number $x$ and tries to find the least positive integer $r$
such that $x^r\equiv 1 \pmod{n}$.
This least $r$ is the {\em order} of $x$ mod $n$, and $n$ can be factored with a high
probability, once $r$ is known.

Shor's algorithm does not yield $r$ directly (in the quantum process).
Instead, another integer $c$ is the actual outcome of the quantum protocol,
from which the right $r$ can
{\em sometimes} be obtained by a classical algorithm.
Let us call the outcome of the classical algorithm $r'$; in at least $O(1/\log\log n)$ 
fraction of the cases, the number $r'$ is the desired $r$, and whether it is the
case or not is checked via a classical algorithm.
Let the probability of a correct result (on an individual computer) be $p_r$.
While the order $r$ (for a given $x$ and $n$) is unique, the result $c$
and the calculated $r'$ are not unique.
Having several good outcomes $c_i$ does not cause a problem
(as noted by in~\cite{CG}), since the quantum computer can perform
a classical
algorithm which calculates $r$ from any of the possible $c_i$.
However, this operation by itself is not sufficient, since many
of the computers (probably, most of them) give an outcome $r'$ which
is not the correct $r$. When expectation values are measured for the $j$th
bit, the correct result $r_j$ happens with small probability $p_r$, and hence
it is obscured by the wrong results $r_j'$.

If the measurement process could distinguish among $2^n$ states of an 
$n$-qubit system (which will require exponential resolution), then one could
read the correct result accurately. However, such an operation
is not permitted, and hence the technique of \cite{CG} is not sufficient.  Another
potential situation, which could also lead to a simple resolution, is if  
the wrong-$r$ results are well
distributed (e.g., totally random); in such a case, on the average these wrong-$r$ results
will cancel out (e.g., average to yield zero) and will not obscure the correct
result. Let us show that this is not always the case, and that the bad results are
not always averaged to zero, and hence the good result sometimes is indeed obscured.

The output $c$ of the quantum process in Shor's algorithm is used to calculate
the order $r$~\cite{Shor}. For this, the integers $d'$ and $r'$ are found such that
\[ \left | \frac{c}{q}-\frac{d'}{r'}\right | \leq \frac{1}{2q}, \]
where $n^2<q\leq 2n^2$, and $q$ is a power of 2. Then the fraction $d'/r'$ is unique. 
The integer $r'$ is the output of the algorithm as the desired order (which is actually
$r$). To continue,
let $\alpha(c)$ be the unique integer such that  $-q/2\leq\alpha(c)\leq q/2$ and
$rc\equiv \alpha(c) \pmod{q}$. One of the possible situations that leads to 
incorrect answer is that the output $c$ of the quantum process satisfies the condition
\[ \left | \frac{c}{q}-\frac{d}{r}\right | \leq \frac{1}{2q}, \]
and $d$ and $r$ {\em are not} relatively prime. Then the answer, instead of $r$, would be
a divisor of $r$. The probability that such event occurs is (see~\cite{Shor}) approximately
$4(r-\phi(r))/(\pi^2 r)$. This probability can be some constant far away from zero.
For example, if $r=2^s3^t$, then $\phi(r)=r/3$ and the probability the algorithm
provides a divisor of $r$ is $\approx 0.135$.

Let us now present a modified factorization protocol that bypasses this ensemble
measurement problem.
The idea is to replace an additional part of the classical protocol, a part
which verifies that $r$ is indeed the order, by a
quantum one. Also, a simple (but crucial)
modification of the protocol is required.
Let the register holding the result ($r$ or $r'$) be called $s_1$.
Let us use an additional register $s_2$ of the same number $\ell$ of qubits as
$s_1$. Let the register $s_2$ be in the state
\begin{equation}
 \mbox{\footnotesize $H|0\rangle\otimes H|0\rangle\otimes\cdots\otimes H|0\rangle=
  \frac{1}{2^{\ell/2}}\sum_{x\in\{0,1\}^\ell}\ket{x} $} ,
\label{random}
\end{equation}
where $H|0\rangle =\frac{1}{\sqrt{2}}(|0\rangle + |1\rangle )$.
Now we augment the quantum factorization algorithm with the following procedure.
When the original factorization algorithm finishes, test the result in the
register $s_1$ to see whether it gives the correct value of the order $r$. 
If the result on the $i$th computer is indeed the order then 
nothing is to be done and the outcome $r$ is kept in $s_1$.
Whenever the result is an incorrect value $r'$, swap the contents of the registers 
$s_1$ and $s_2$ so the outcome $r'$ is replaced by the state 
$H|0\rangle\otimes\cdots \otimes H|0\rangle$
which yields a completely randomized outcome once it is measured.
Now, a measurement of the $j$th bit on $s_1$ will give the correct result
if the string holds the state $r$ or it yields zero (on average) if the
string
originally contained the wrong result $r'$.

Although the strength of the good signal may be small, there are enough
computers running in parallel to read it since in the worst case, 
it is only logarithmically small.

\subsection{Algorithms for NP functions}

Technique used in the previous section for Shor's algorithm can easily be
generalized to any quantum algorithm that computes an NP function. By an
{\em NP function} we mean a function whose graph is in the class P. 
More specifically, a function $f:\Sigma^*\longrightarrow\Sigma^*$, for
some
alphabet $\Sigma$, such that there is a polynomial--time Turing machine that
for $x,y\in\Sigma^*$ decides whether $f(x)=y$ or not.

\subsection{The search algorithm}

Certain search operations in a database can be done more efficiently on a quantum computer 
than on a classical computer \cite{Grover}. Here the search means to find
some item $x$ in the database such that $x$ satisfies some predefined condition
$T$; i.e, we are looking for the solutions of $T(x)=1$. The analysis of
\cite{BBHT} shows that if the size of the database is $N$ and the number of
solutions are $t$, Grover's algorithm, with high probability, can find a solution
in time $(O\sqrt{N/t})$. When there is only one solution,
this algorithm yields the desired result also on an ensemble computer.

However, when several (say $t\geq 2$) different items satisfy the required
condition, the protocol will randomly yield one of them. Therefore, in this case 
the algorithm is not suitable for ensemble computation. We show here how this algorithm
can be modified such that ensemble computation still provides a correct solution with high 
probability.

We assume $t$, the number of solutions, is {\em known} and {\em constant} (the general
case will be studied in the next section). We first consider the case $t=2$.
When processed on an ensemble--measurement  computer, only expectation 
values are 
obtained, and the two outcomes partially obscure 
each other to yield zero (as the 
average expected value) for $j$th bit of the answer if the $j$th bits
of the two solutions are different.

To solve this problem we suggest to hold several (say $m$)
computers in one molecule.
After each computer in the molecule finishes Grover's algorithm,
the procedure is continued by sorting the outputs of different computers in an
increasing order. Finally let the algorithm contain a step where the first and the 
last results are compared, and if they are equal then both are replaced by 
a randomized data (\ref{random}), as in the modified Shor's algorithm.
Once the first and last computers hold different outcomes, we are promised
that the small solution is always the first, and that
the large solution is always the last. Thus, we can obtain both solutions.

The probability that the first and the last solutions are the same is
$\frac{1}{2^m}$, so the final outcome is obtained with probability exponentially
close to one. Even without applying the randomization to the bad outcomes,
the expected outcomes are still readable. 

When $t>2$, we apply the same procedure (without randomization to the bad outcomes).
We still reorder the solutions so that the minimal solution is in the first position.
However, we might obtain different minimal solutions for different molecules.
The probability of failing to obtain the global minimum solution in the first 
position is $(1-\frac{1}{t})^m$, and as long as it is small (say less than $e^{-\lambda}$, 
which holds if $m > \lambda t$) the protocol can work properly. Note that this 
modified algorithm still works in time $O(\sqrt{N/t})$.

Only the smallest and largest solutions can be obtained by the above method.
If one needs the other solutions, these can easily be obtained via similar methods,
once some solutions are already known.

\subsection{Search algorithm: the case of unknown number of solutions}

Now we consider the most general case. Here we do not assume any condition on $t$,
the number of solutions; it can be known or unknown, large or even zero. Our
method is based on a binary search. We also utilize the following fact established
in \cite{BBHT}: Let $\cal B$ be a database of size $M$;
then the search algorithm, with high probability, starting with the input 
$\frac{1}{\sqrt{M}}\sum_{x\in{\cal B}}\ket{x}$ in time $O(\sqrt{M})$ can determine 
whether there is any solutions in $\cal B$ or not.

Without loss of generality, we can assume that 
the database is represented as the members of the unit cube $V=\{0,1\}^n$. So 
$N=2^n$. For any string $\alpha=(\alpha_1,\ldots,\alpha_k)\in\{0,1\}^k$, let 
$V_\alpha$ be the subset of $V$ consisting of all strings 
$(\alpha_1,\ldots,\alpha_k,x_{k+1},\ldots,x_n)$; i.e., $V_\alpha$ contains all
strings in $V$ that start with $\alpha$. Thus $|V_\alpha|=2^{n-k}$.

Our algorithm first checks whether there is a solution or not. If there is no solution
then it stops. Otherwise it runs in $n$ stages. The output of the stage $j$ is a 
database ${\cal B}_j$ of size $2^{n-j}$ which contains a solution.
At the end ${\cal B}_n=\{\xi\}$, where $\xi$ is a solution.
The algorithm starts with the database ${\cal B}_0=V$. It checks whether there is
any solution in $V_0$. If there is a solution then ${\cal B}_1=V_{0}$,
otherwise ${\cal B}_1=V_{1}$. In a general stage $j+1$, the input is of the form
${\cal B}_j=V_{\alpha_j}$ where $\alpha_j\in\{0,1\}^j$, and there is a solution in
${\cal B}_j$. Then the algorithm checks whether there is a solution in $V_{\alpha_j0}$,
if so then the output of this stage is ${\cal B}_{j+1}=V_{\alpha_j0}$, otherwise the
output is ${\cal B}_{j+1}=V_{\alpha_j1}$. This completes the description of our
search algorithm. It is easy to check that this algorithm always provides the 
{\em first} solution in the lexicographic order. So we have presented a quantum search 
algorithm that always gives a unique output, no matter how many solutions are there.
This is an algorithm which can be implemented on an ensemble--measurement computer. Note that
the running time of this algorithm is
\[O\left(\sqrt{2^n}+\sqrt{2^{n-1}}+\cdots+\sqrt{2}\,\right)=O\left(\sqrt{N}\,\right).\]

\section{Error--recovery in the error--correction process}
Standard error correction can be viewed as a computation with
more than one good answer, and thus belongs to Case (1) discussed
in Section 2.  In this case, the syndrome of the error is not
unique.  In the standard prescription, measurement is used to collapse
the ancilla qubits containing the error information.  Then these syndrome bits 
are processed by a classical reversible algorithm to determine the errors, and 
a unitary operation to correct the error is applied to the data
qubits by the output bits of the classical algorithm.  In the measurement-free case, 
the ancilla qubits need not be measured, and the classical subroutine (following
the measurement) could be incorporated into the original quantum algorithm.  

One can easily verify that the above-mentioned classical subroutine (that processes
the ancilla qubits) needs to use Toffoli gates. Using the techniques of Section 3 one could
implement a quantum Toffoli gate without measurements, and hence, there is no fundamental
problem in having a single quantum code for the measurement-free circuit. However,
implementing a quantum Toffoli gate fault tolerantly and without measurement
is an involved process. Fortunately, the techniques of Section 3 can be also applied so that
the classical subroutine is carried out on a classical code. The state
of the ancilla qubits can be first copied onto a classical repetition code using the
$\cal N$ gate.  Now classical reversible computation can be performed on the repetition
code and then a control operation can be performed on the quantum data to correct for the
errors.  Since phase errors from the classical subcircuit will not propagate to 
the quantum data, using repetition
codes to correct for any bit errors in the subcircuit is sufficient. 
This technique thus allows one to fault-tolerantly
replace quantum Toffoli gates by classical ones in the error recovery process.

\section{concluding Remarks}

To summarize, we showed that running algorithms on bulk (ensemble)
computers is not always obvious.
We modified various important algorithms so that they can run on ensemble computers.

More work is required in order to run algorithms without measurement with only
near-neighbor interactions, and more work is required to solve the addressing and
scaling problems.

We are thankful to Dorit Aharonov for many helpful remarks.

{\bf APPENDIX A}

As mentioned in the introduction, in NMR computing, each molecule is used as
a computer, and
different qubits in one computer are spins of different nuclei.
Many identical molecules are used (a macroscopic number) in parallel. Moreover, 
the state of the qubits is initially a thermal mixture.

There are three main problems with the current proposals for NMR computers
~\cite{Warren,DiVince}: the ensemble--measurement problem, the
addressing problem, and the pseudo-pure-state scaling problem.
Unless these problems can be solved or mitigated, it is widely believed
that NMR computing will not be very useful as a future computing 
device. As we demonstrate in this paper, the ensemble-measurement problem can
be addressed successfully. While the other two problems are challenging,
we argue in the following paragraphs that recent advances do hold the
promise of 
mitigating their effects, and that further research is required before one can
conclude whether NMR/ensemble quantum computing can indeed be scaled up to
perform practical
quantum computation.

{\em The addressing problem:} The individual qubits in an NMR computing system
cannot be accessed by a laser directed only to it, and hence different level
separation is usually used for each of the qubits. For $n$ qubits (with
only near--neighbor interactions) there is a need for $O(n)$ different
laser frequencies, and off--resonance effects become
non--negligible.  A solution to this
problem was suggested in~\cite{Lloyd}, where a chain of three different types of
qubits, arranged in the form of $ABCABCABC\ldots ABC$ is used. 
In this chain one (and only one) of the qubits, say of type $B$,
is replaced by a pointer--qubit of a fourth type $D$.  Now, with only five
swaps
operations: {\sf swap}($AB$), {\sf swap}($CA$), {\sf swap}($BC$),
{\sf swap}($AD$), {\sf swap}($DC$), and with
three operations on the pointer and its neighborhood: individual qubit
rotations $R(D)$, and two--qubit operations $U(AD)$ and $U(CA)$,
algorithms can run with only a polynomial slowdown. Thus, in principle,
universal quantum
computation can be performed on an NMR system with only a constant
number of laser frequencies.

{\em The pseudo-pure-state scaling problem:} The state of the qubits in an NMR
computer is highly mixed. It is a thermal mixture so that the qubits 
are in a state that is $|0\rangle$ with probability $\frac{1+ \epsilon}{2}$ and
in a state which is $|1\rangle$
with probability $\frac{1-\epsilon}{2}$, where $\epsilon$ is a function of the
temperature and the applied strong magnetic field.
For the quantum computation model, however, it is assumed that initially all
its qubits
are in a known state, which, without loss of generality, is assumed to be
the state
$|0\rangle$. In the existing literature (and current experiments), 
a novel purification technique was used, which creates a ``pseudo-pure-state'',
that is, a state which can be written as a mixture of the identity and a pure state.
Then, the algorithmic steps are performed on the ``pseudo--pure state''. While
this 
ingenious technique allows one to perform entanglement manipulation and
demonstrate quantum
algorithms involving a few qubits,  it has an inherent limitation. In
particular,  
there is an information loss in the process of mixing (via a non--unitary
operation) of
all eigenstates except the state $|000...0\rangle$ (see, for example,
\cite{CG} for detailed
explanations), leading to an exponential decrease in signal--to--noise ratio 
with the increase in the number of qubits. Hence, the current pseudo--pure state
approaches cannot be scaled up, and thus they lose any potential advantage
over classical computers.
It is worth observing, however, that the exponential loss of signal is {\em
an artefact of  the existing
pseudo--pure state approaches}, and is {\em not inherent to NMR or ensemble
quantum computing}.
For example, a simple information--theoretic analysis suggests that
$k = O (n \epsilon^2)$ pure qubits can be distilled from $n$ thermal
qubits, which are highly mixed. 
This idea was analyzed further in \cite{SV}, where an algorithm for extracting
$O (n \epsilon^2)$ pure qubits from a thermal mixture of $n$ qubits was
suggested.
While the solution of~\cite{SV} is good only when $n$ is large compared to
$\epsilon^2$, it clearly proves the point that methods for creating much
better pseudo--pure states probably exist, and that the scaling problem should
certainly not discourage scientists from pursuing ensemble quantum computation.
We are currently working on this problem and the initial results are very
promising.


\begin{thebibliography}{99}

\bibitem{Dorit}
D. Aharonov and M. Ben-Or, 
``Fault-Tolerant Quantum Computation with Constant Error,'' {\em Proc. of
the 29th Annual ACM Symposium on Theory of Computing (STOC)}, pp. 46-55, 1997.

\bibitem{Dorit2}
D. Aharonov and M. Ben--Or, ``Fault-Tolerant Quantum Computation With Constant Error
Rate'', the journal version of~\cite{Dorit},
Los-Alamos archive: Quant-ph/9906129.

\bibitem{BBHT}
M. Boyer, G. Brassard, P. Hoyer and A. Tapp, ``Tight bounds on quantum searching,''
 {\em Fortschritte der Physik}, 46(1998), pp. 493--505. 

\bibitem{sigma_z}
P. O. Boykin, T. Mor, M. Pulver, V. Roychowdhury, and F. Vatan, 
``On universal and fault--tolerant quantum computation,'',
Los-Alamos archive Quant-ph/9906054.
To appear in {\em Proc. 40th
IEEE Ann. Symposium on Foundations of Computer Science (FOCS)}, 1999.

\bibitem{Teleport1} G. Brassard, S. Braunstein, and R. Cleve,
``Teleportation as a quantum computation'', {\em Physica D}, 120(1998), pp. 43--47. 

\bibitem{QEC} D. Cory, M. Price, W. Mass, E. Knill, R. Laflamme,
W.~Zurek, T.~Havel, and S.~Somaroo, 
``Experimental quantum error correction'', {\em Physical Review Letters},
81(1998), pp. 2152-2155.

\bibitem{DiVince} D. DiVincenzo, ``Real and realistic quantum computation'',
{\em Nature}, 393(1998), pp. 113--114.

\bibitem{DS} 
D. DiVincenzo and P. Shor, ``Fault-tolerant error correction with efficient quantum 
codes,'' {\em Physical Review Letters}, 77(1996), pp. 3260--3263.

\bibitem{CG} 
N. A. Gershenfeld and I. L. Chuang, ``Bulk spin-resonance quantum computation,''
{\em Science}, 275(1997), pp. 350--356.

\bibitem{Cory} D. G. Cory, A. F. Fahmy, and T. F. Havel, ``Ensemble quantum computing
by nuclear magnetic resonance spectroscopy,'' in {\em Proc. Natl. Acad. Sci.} 
94(1997), pp. 1634--1639.

\bibitem{Grover} L. Grover, ``A fast quantum mechanical algorithm for
database search,'' in {\em Proceedings of 28th ACM Symposium on Theory
of Computing}, pp. 212--219, 1996. 

\bibitem{Mosca} J. A. Jones, M. Mosca, and R. H. Hansen, 
``Implementation of a quantum search algorithm on a quantum computer'',
{\em Nature}, 393(1998), pp. 344--346.

\bibitem{Kitaev}  A. Kitaev, ``Quantum Computations: Algorithms
and Error Correction'', {\em Russian Math. Surveys} 52(1997), pp. 1191-1249.

\bibitem{KLZ2}
E. Knill, R. Laflamme, and W. H. Zurek, 
``Accuracy Threshold for Quantum Computation'',
Los Alamos archive: quant-ph/9610011.

\bibitem{KLZ1}
E. Knill, R. Laflamme, and W. H. Zurek, 
``Resilient quantum computation: error models and thresholds,''  
{\em Proceedings of the Royal Society of London, Series A}, 454(1998), pp. 365-384.

\bibitem{Lloyd} 
S. Lloyd, ``Universal quantum simulators,'' {\em Science}, 273(1996), pp. 1073-1078.

\bibitem{Teleport2} M. A. Nielsen, E. Knill, and R. Laflamme, 
``Complete quantum teleportation using nuclear magnetic resonance'',
{\em Nature}, 396(1998), pp. 52--55.

\bibitem{Peres}
A. Peres, ``Quantum disentanglement and computation,''
{\em Superlattices and Microstructures}, 23(1998), pp. 373--379.

\bibitem{Preskill}
J. Preskill, ``Reliable quantum computers,'' {\em Proc. of the Royal Society of 
London, Ser. A},  454(1998), pp. 385--410.

\bibitem{SV} L. J. Schulman and U. Vazirani, ``Scalable NMR quantum computing,''
  LANL e--print, quant--ph/9804060, 1998.

\bibitem{Shor} P. Shor, ``Polynomial--time algorithms for prime
factorization and discrete logarithms on a quantum computer,'' {\em SIAM
J. Computing}, 26(1997), pp. 1484--1509. 

\bibitem{Shor_fault}
P. Shor, ``Fault--tolerant quantum computation,'' in {\em Proc. 37th
IEEE Ann. Symposium on Foundations of Computer Science}, pp. 56--65, 1996.

\bibitem{Warren} W. S. Warren, ``The usefulness of NMR quantum computing'',
{\em Science}, 277(1997), pp. 1688--1689.

\end{thebibliography}
\end{document}